\RequirePackage{fix-cm}
\documentclass[smallextended]{svjour3}       
\smartqed 
\usepackage{graphicx,amsmath,amssymb,wrapfig}
\begin{document}

\title{Hadron tomography and its application to gravitational radii of hadrons
\thanks{This work was supported by Japan Society for the Promotion of Science (JSPS)
Grants-in-Aid for Scientific Research (KAKENHI) Grant Number JP25105010.}
}
\author{S.~Kumano\and Qin-Tao~Song\and O.~V.~Teryaev}
\institute{S.~Kumano\at
           KEK Theory Center, Institute of Particle and Nuclear Studies, KEK
           and Department of Particle and Nuclear Physics,
           Graduate University for Advanced Studies (SOKENDAI),\\
           Ooho 1-1, Tsukuba, Ibaraki, 305-0801, Japan\\
             J-PARC Branch, KEK Theory Center,
             Institute of Particle and Nuclear Studies, KEK
             and Theory Group, Particle and Nuclear Physics Division, J-PARC Center,
             203-1, Shirakata, Tokai, Ibaraki, 319-1106, Japan
           \and
           Qin-Tao~Song\at
           Department of Particle and Nuclear Physics,
           Graduate University for Advanced Studies (SOKENDAI)
           and KEK Theory Center, Institute of Particle and Nuclear Studies, KEK,
           Ooho 1-1, Tsukuba, Ibaraki, 305-0801, Japan
           \and
           O.~V.~Teryaev\at
           Bogoliubov Laboratory of Theoretical Physics,
           Joint Institute for Nuclear Research,\\
           141980 Dubna, Russia
}

\date{Received: January, 2018 / Accepted: }
\maketitle

\begin{abstract}
Hadron tomography has been investigated by three-dimensional structure functions,
such as generalized parton distributions (GPDs) and generalized
distribution amplitudes (GDAs). The GDAs are $s$-$t$ crossed quantities
of the GPDs, and both functions probe gravitational form factors for hadrons.
We determined the pion GDAs by analyzing Belle data on the differential
cross section for the two-photon process $\gamma^* \gamma \to \pi^0 \pi^0$.
From the determined GDAs, we calculated timelike gravitational form factors
of the pion and they were converted to the spacelike form factors
by using the dispersion relation. These gravitational form factors 
$\Theta_1$ and $\Theta_2$ indicate mechanical (pressure, shear force)
and gravitational-mass (or energy) distributions, respectively.
Then, gravitational radii are calculated for the pion from the form factors,
and they are compared with the pion charge radius. We explain that
the new field of gravitational physics can be developed in the microscopic
level of quarks and gluons.
\end{abstract}

\section{Introduction}
\label{intro}

Three dimensional (3D) tomography has been investigated recently 
for the nucleon by using generalized parton distributions (GPDs) 
and transverse-momentum-dependent parton distributions (TMDs)
from experimental measurements on
the deeply virtual Compton scattering (DVCS) and
semi-inclusive deep inelastic scattering, respectively \cite{gpds-gdas}.
Generalized distribution amplitudes (GDAs) are $s$-$t$ crossed
quantities of the GPDs and they can be investigated by the two-photon
process $\gamma^* \gamma \to h \bar h$ to produce a hadron ($h$) pair.
The major reasons for studying the 3D tomography is 
(1) to find the origin of nucleon spin including orbital angular momentum 
contributions \cite{gpds-gdas}, 
(2) to find internal structure of exotic hadron candidates \cite{gdas-kk-2014}, 
(3) to investigate gravitational properties of hadrons \cite{gdas-kst-2017}.

In 2016, the Belle collaboration published the cross-section data
on $\gamma^* \gamma \to \pi^0 \pi^0$ \cite{Masuda:2015yoh}, 
so that it became possible to discuss the GDAs in comparison 
with actual experimental measurements.
We determined the pion GDAs by analyzing the Belle data,
and then gravitational form factors were evaluated for the pion
from the obtained GDAs \cite{gdas-kst-2017}. 
We discuss these results in this report.
First, the definitions of the GPDs and GDAs are introduced
in Sec.\,\ref{3d-sfs}, gravitational form factors of the pion
are explained in Sec.\,\ref{grav-ffs},
and the differential cross section 
of $\gamma^* \gamma \to \pi^0 \pi^0$ is expressed by the GDAs
in Sec.\,\ref{cross}. Our analysis results are discussed
for the GDAs and gravitational form factors in Sec.\,\ref{results}.

\section{Theoretical formalism}
\label{formalism}

\subsection{Three-dimensional structure functions}
\label{3d-sfs}

\begin{figure}[t!]
\begin{minipage}{\textwidth}
\begin{tabular}{lcl}
\begin{minipage}[c]{0.45\textwidth}
\hspace{0.40cm}
    \includegraphics[width=5.0cm]{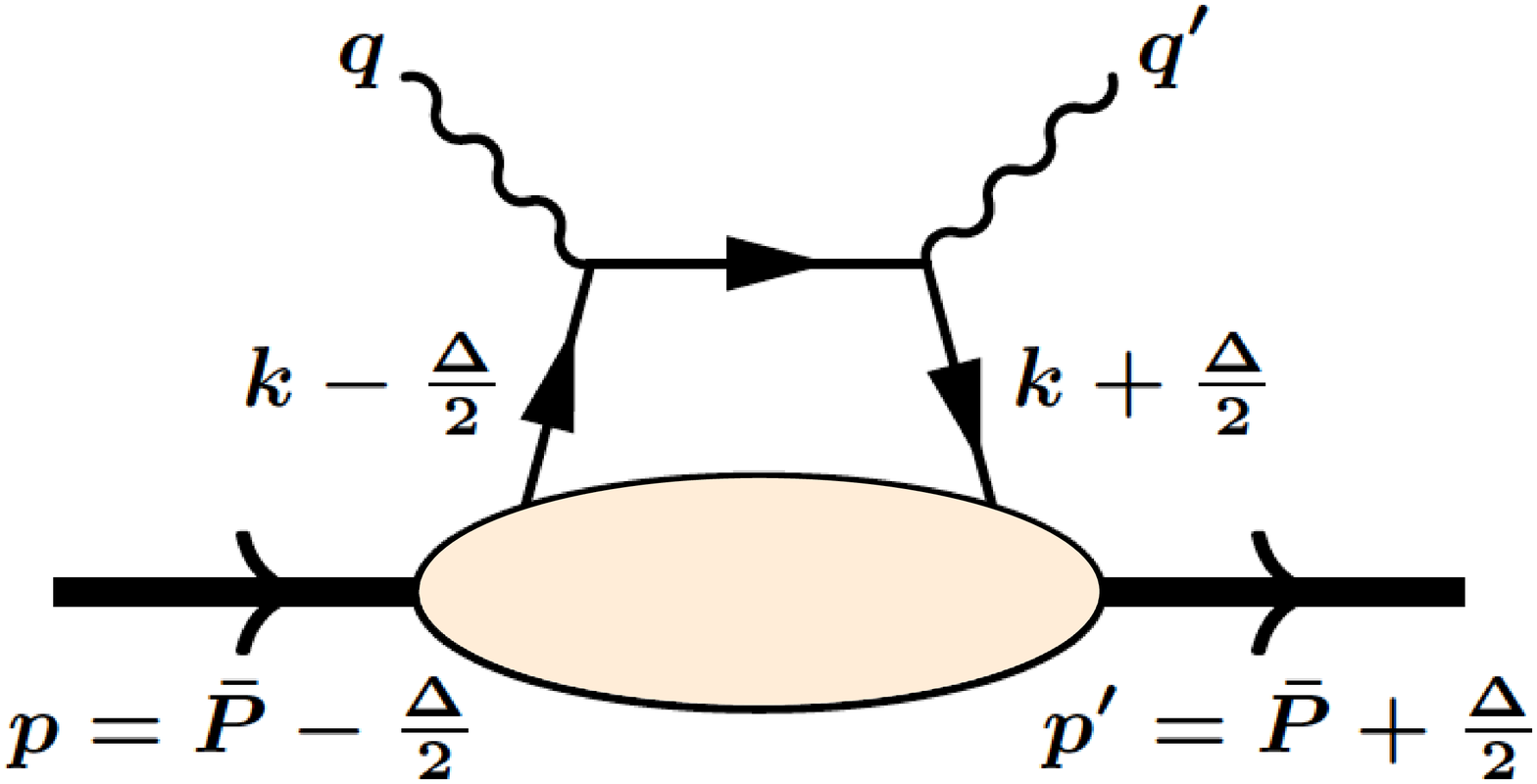}
\end{minipage} 
\hspace{1.50cm}
\begin{minipage}[c]{0.45\textwidth}
    \includegraphics[width=4.2cm]{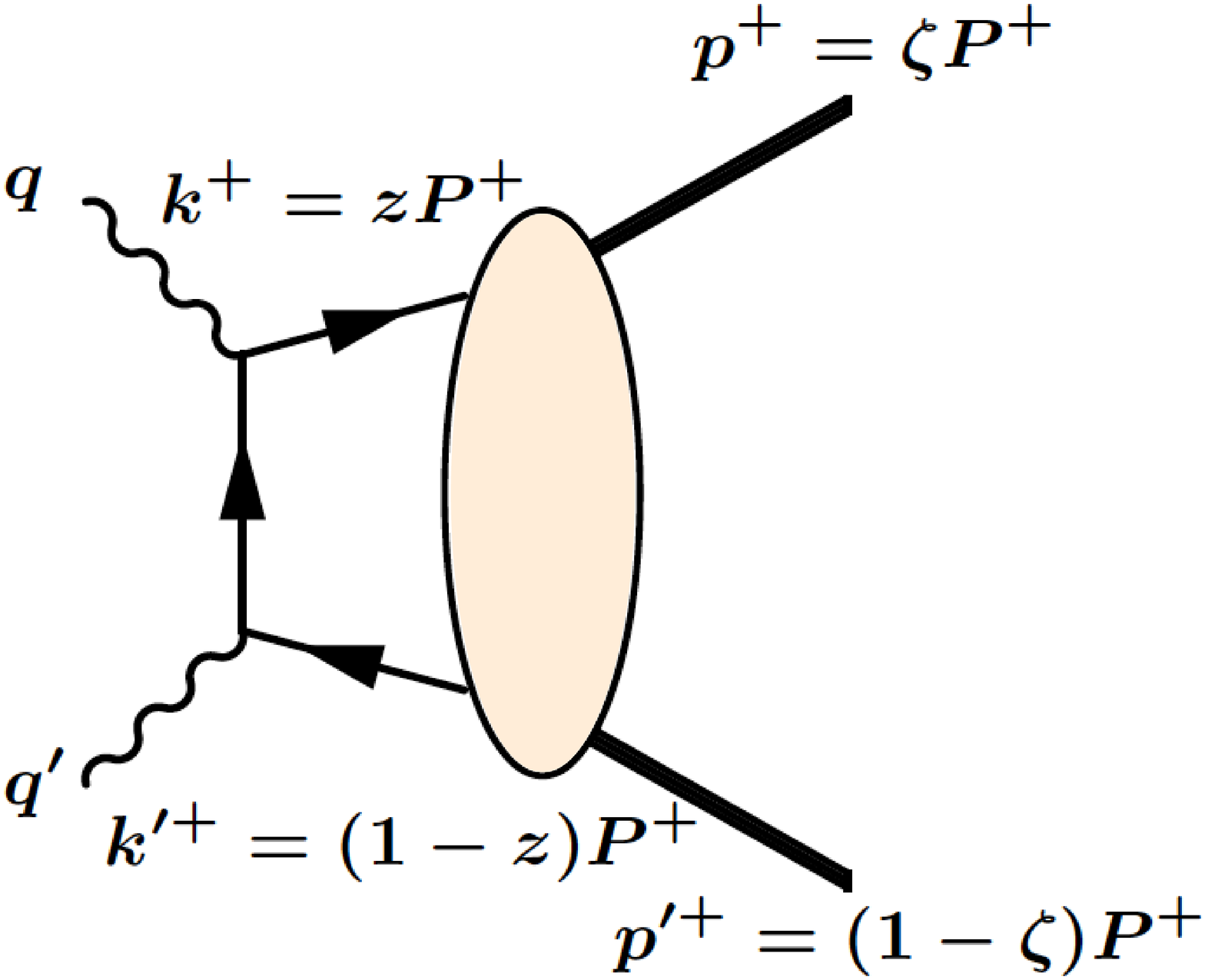}
\end{minipage}
\end{tabular}
\end{minipage}
\ \vspace{-0.00cm}\hspace{2.9cm}
$(a)$ \hspace{5.2cm} $(b)$
\vspace{-0.1cm}
\caption{\ $(a)$ Virtual Compton process for GPDs.
\ \ \ \ \ \ \ 
$(b)$ Two-photon process for GDAs.}
\label{fig:gpd-fig}
\vspace{-0.3cm}
\end{figure}

The 3D structure functions, the GPDs and GDAs, for the hadron $h$
are measured by the deeply virtual Compton scattering $\gamma^* h \to \gamma h$
and two-photon process $\gamma^* \gamma \to h \bar h$, respectively,
as shown in Fig.\,\ref{fig:gpd-fig}.
In this article, we explain the GPDs and GDAs for the pion.
The quark GPDs $H_q^{\, \pi^0}$ for $\pi^0$ 
are defined by off-forward matrix elements
of quark operators with a lightcone separation, and 
the quark GDAs $\Phi_q^{\, \pi^0 \pi^0}$ are defined in the same way 
by the matrix element between the vacuum and the hadron pair
\cite{gpds-gdas}:
\begin{align}
\! \! \! 
H_q^{\, \pi^0} (x,\xi,t)
& = \! \int\frac{d y^-}{4\pi} \, e^{i x \bar P^+ y^-}
\! \left< \pi^0 (p') \left| 
\overline{q}(-y/2) \gamma^+ q(y/2) 
 \right| \pi^0 (p) \right> \Big |_{y^+ = \vec y_\perp =0}, 
\label{eqn:gpd-pi}
\end{align}
\begin{align}
\Phi_q^{\, \pi^0 \pi^0} (z,\zeta,W^2) 
& = \int \frac{d y^-}{2\pi}\, e^{i (2z-1)\, P^+ y^- /2}
\nonumber \\
& \ \ \ \ \ \ \ \ \ \ 
 \times
  \langle \, \pi^0 (p) \, \pi^0 (p') \, | \, 
 \overline{q}(-y/2) \gamma^+ q(y/2) 
  \, | \, 0 \, \rangle \Big |_{y^+=\vec y_\perp =0} \, .
\label{eqn:gda-pi}
\end{align}
Here, the link variables to satisfy the color gauge invariance
are not explicitly written.

Using the initial and final pion (photon) momenta $p$ and $p'$
($q$ and $q'$) in Fig.\,\ref{fig:gpd-fig}$(a)$, 
we define average momenta ($\bar P$, $\bar q$)
and momentum transfer $\Delta$ as
$\bar P = (p+p')/2$, $\bar q = (q+q')/2$, 
and $\Delta = p'-p = q-q'$.
The GPDs are expressed by three kinematical variables,
the Bjorken variable $x$, the skewness parameter $\xi$,
and the momentum-transfer squared $t$ as
\begin{align}
x = \frac{Q^2}{2p \cdot q} , \ \ \ \ 
\xi = \frac{\bar Q^2}{2 \bar P \cdot \bar q} , \ \ \ \ 
t = \Delta^2 , 
\end{align}
where $Q^2=-q^2$ and $\bar Q^2=-\bar q^2$.
The GDAs are expressed by three different variables,
the momentum fractions $z$ and $\zeta$ in Fig.\,\ref{fig:gpd-fig}$(b)$
and the invariant-mass squared $W^2$ as
\begin{align}
z = \frac{k \cdot q'}{P \cdot q'} = \frac{k^+}{P^+} , \ \ \ 
\zeta & = \frac{p \cdot q'}{P \cdot q'}
      = \frac{p^+}{P^+} = \frac{1+\beta \cos\theta}{2} ,
\nonumber \\
W^2 & = (p+p')^2 = (q+q')^2 = s ,
\end{align}
where $P$ is given by $P=p+p'$,
$a^+$ indicates the lightcone quantity 
$a^+ = (a^0 + a^3 )/ \sqrt{2}$,
$\beta$ is defined by
$\beta =|\vec p \,|/p^0 = \sqrt{1-4m_\pi^2/W^2}$,
and $\theta$ is the scattering angle in the center-of-mass frame
of final pions.

The DVCS process is factorized into the hard perturbative QCD part
and the GPDs as shown in Fig.\,\ref{fig:gpd-fig}$(a)$, 
if the kinematical condition $Q^2 \gg |t|,\ \Lambda_{\text{QCD}}^2$
where $\Lambda_{\text{QCD}}$ is the QCD scale parameter, is satisfied. 
In the same way, the two-photon process is factorized with the GDAs
as shown in Fig.\,\ref{fig:gpd-fig}$(b)$ if the condition 
$Q^2 \gg W^2,\ \Lambda_{\text{QCD}}^2$ is met.
Then, the $\gamma^* \gamma \to \pi^0 \pi^0$ cross section is 
expressed by the GDAs, which can be determined by analyzing
the Belle data.

\subsection{Gravitational form factors}
\label{grav-ffs}

The GPDs and GDAs contain information on gravitational sources
in the quark and gluon level. In order to show it, we take the $n$-th
moment of the bilocal operator defining the GDAs in 
Eq.\,(\ref{eqn:gda-pi}) as
\begin{align}
2(P^+/2)^{n} \int_0^1 dz (2z-1)^{n-1} 
& \int\frac{d y^-}{2\pi}e^{i (2z-1) P^+ y^- /2}
\overline{q}(-y/2) \gamma^+ q(y/2) \Big |_{y^+ = \vec y_\perp =0}
\nonumber \\
& = \overline q (0) \gamma^+ \!
 \left ( i \overleftrightarrow \partial^+  \right )^{n-1} 
 q(0) .
\label{eqn:tensor-int}
\end{align} 
\begin{wrapfigure}[10]{r}{0.50\textwidth}
   \vspace{-0.3cm}
   \begin{center}
     \includegraphics[width=6.0cm]{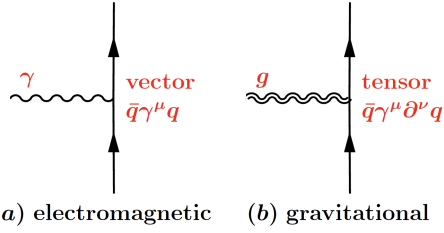}
   \end{center}
\vspace{-0.40cm}
\caption{Electromagnetic and gravitational form factors.}
\label{fig:electro-grav}
\vspace{-0.7cm}
\end{wrapfigure}
\noindent
This equation indicates that the operator is the usual vector
type $\bar q \gamma^\mu q$ for $n=1$, so that the electromagnetic 
form factor is probed for the pion as shown in 
Fig.\,(\ref{fig:electro-grav})$(a)$.
However, the $n=2$ term indicates the energy-momentum tensor 
for quarks as shown in Fig.\,(\ref{fig:electro-grav})$(b)$,
and its form factors can be obtained by studying the GPDs and GDAs.
In fact, the second moment of the GDAs is given by the matrix element
of the quark energy-momentum tensor $T_q^{\mu\nu}$ as
\vspace{-0.00cm}
\begin{align}
\int_0^1  dz \, (2z -1) \, 
\Phi_q^{\pi^0 \pi^0} (z,\,\zeta,\,W^2) 
= \frac{2}{(P^+)^2} \langle \, \pi^0 (p) \, \pi^0 (p') \, | \, T_q^{++} (0) \,
       | \, 0 \, \rangle .
\label{eqn:integral-over-z}
\end{align}
The quark energy-momentum tensor is generally defined by
$ T_q^{\,\mu\nu} (x) = \overline q (x) \, \gamma^{\,(\,\mu} 
   i \overleftrightarrow D^{\nu)} \, q (x)$
with the covariant derivative 
$D^\mu = \partial^{\,\mu} -ig \lambda^a A^{a,\mu}/2$.
Here $g$ is the QCD coupling constant and 
$\lambda^a$ is the SU(3) Gell-Mann matrix.
The gluon energy-momentum tensor is also defined in the same way.
By the matrix element of the energy-momentum tensor,
the timelike gravitational form factors $\Theta_1$ and $\Theta_2$
of the pion are defined as
\vspace{-0.00cm}
\begin{align}
\! \! \! \!
\langle \, \pi^0 (p) \, \pi^0 (p') \, | \, T_q^{\mu\nu} (0) \, | \, 0 \, \rangle 
= \frac{1}{2} 
  \left [ \, \left ( s \, g^{\,\mu\nu} -P^{\,\mu} P^\nu \right ) \, \Theta_{1, q} (s)
                + \Delta^\mu \Delta^\nu \,  \Theta_{2, q} (s) \,
  \right ] ,
\label{eqn:emt-ffs-timelike-0}
\end{align}
where $P$ and $\Delta$ are $P=p+p'$ and $\Delta=p'-p$.
Therefore, once the GDAs are determined, we can calculate these form factors
by using Eqs.\,(\ref{eqn:integral-over-z}) and (\ref{eqn:emt-ffs-timelike-0}).
We know that the energy-momentum tensors of quarks and gluons are the sources
of gravity, so that they are called gravitational form factors.
The gravitational interactions are generally too weak to be investigated in
particle scattering experiments; however, the GPDs and GDAs provide
a way to access them from the microscopic quark and gluon level.

\vspace{-0.10cm}
\subsection{Cross section for $\gamma^* \gamma \to \pi^0 \pi^0$ and GDAs}
\label{cross}
\vspace{-0.10cm}

The cross section for the two-photon process 
$\gamma^* \gamma \to \pi^0 \pi^0$ is given by the matrix element 
${\cal M}$ as
\vspace{-0.00cm}
\begin{align}
d\sigma = \frac{1}{4 q\cdot q'}
\underset{\lambda, \lambda'}{\overline\sum}
| {\cal M} (\gamma^* \gamma \to \pi^0 \pi^0 ) |^2 
\frac{d^3 p}{(2\pi)^3 \, 2E_p}  \frac{d^3 p'}{(2\pi)^3 \, 2E_{p'}} 
(2\pi)^4 \delta^4(q \! + \! q' \! - \! p \! -\! p') . 
\nonumber \\[-0.40cm]
\label{eqn:cross-section}
\end{align}
The matrix element is expressed by the hadron tensor ${\cal T}_{\mu\nu}$
and the photon polarization vector $\epsilon^\mu$ as
$ i {\cal M} (\gamma^* \gamma \to \pi^0 \pi^0 ) 
   = \epsilon^\mu(\lambda) \, \epsilon^\nu(\lambda') \, {\cal T}_{\mu\nu} $
with
\begin{align}
{\cal T}_{\mu \nu } 
& = i \int d^4 y {e^{ - iq \cdot y}} 
\left\langle \pi^0 (p) \pi^0 (p') \left| 
{TJ_\mu ^{em}(y)J_\nu ^{em}(0)} \right|0 \right\rangle 
\nonumber \\[-0.10cm]
& =  - g_{T}^{\, \mu \nu}{e^2} 
\sum\limits_q \frac{{e_q^2}}{2} \!
\int_0^1 \! \! {dz} \frac{{2z - 1}}{{z(1 - z)}}
\Phi_q^{\pi^0 \pi^0}(z,\zeta ,{W^2}) ,
\label{eqn:matrix}
\end{align}
where
$ g_{T}^{\, \mu \nu} = -1 $  for $\mu=\nu=1, \ 2$ and
$ g_{T}^{\, \mu \nu} = 0 $  for $\mu$, $\nu=\,$others.
Then, defining the helicity amplitude $A_{i j}$ as
$ A_{i j}  = \varepsilon _\mu ^{( i )}(q) \, \varepsilon _\nu ^{( j )}(q') \,
{{\cal T}^{\mu \nu }} /e^2$ ($i=-,\ 0, \ + \, ; \ j=-,\ + \, $), 
we obtain the cross section expressed by the GDAs as
\begin{align}
\frac{d\sigma}{d(\cos \theta)}
& = \frac{\pi \alpha^2}{4(Q^2+s)}
    \sqrt{1-\frac{4m_\pi^2}{s}} \, |A_{++}|^2  ,
\nonumber \\
A_{++}
& =\sum_q \frac{e_q^2}{2} \int^1_0 dz \frac{2z-1}{z(1-z)} 
   \Phi_q^{\pi^0 \pi^0} (z, \xi, W^2) .
\label{eqn:cross2}
\end{align}
In our analysis, higher-order effects of $\alpha_s$ and higher-twist terms
are neglected. The gluon GDA contributes to the cross section as 
a higher-order term, so that they are not studied in this work.

\section{Results for GDAs and gravitational form factors for pion}
\label{results}

The GDAs are expressed by a number of parameters, which are determined
by a $\chi^2$ analysis of the Belle cross-section data on 
$\gamma^* \gamma \to \pi^0 \pi^0$.
The possible isospin and angular momentum states for the final two pions
are $I=0$ and $L=\text{even numbers (0, 2, $\cdots$)}$.
We only consider the lowest Gegenbauer polynomial $n=1$ in setting up
the $z$-dependent functional form of the asymptotic GDAs \cite{gdas-kst-2017}. 
Then, the possible states are $L=0$ (S wave) and 2 (D wave).
The GDAs are expressed by the addition of S- and D-wave contributions
$\widetilde B_{10}(W^2)$ and $\widetilde B_{12}(W^2)$ as
\begin{align}
\! \! \!
\Phi_q^{\pi ^0 \pi^0} (z, \zeta, W^2) = 
           N_\alpha z^\alpha(1-z)^\alpha (2z-1) \,
 [\widetilde B_{10}(W^2) + \widetilde B_{12}(W^2) P_2(\cos \theta)] ,
\label{eqn:gda-parametrization}
\end{align}
where $P_2(\cos \theta)$ is the Legendre polynomial.
In the scaling limit $Q^2 \to \infty$, the $z$ dependence is given 
$z (1-z) (2z-1)$ for the $\pi^0$ GDAs. In the above function,
the parameter $\alpha$ is assigned for its functional variation.

There are contributions from the GDA continuum and resonances
to the functions $\widetilde B_{nl}(W^2)$:
$\widetilde B_{10}(W^2)= \text{continuum} + \text{resonance ($f_0$)}$,
$\widetilde B_{12}(W^2) = \\ \text{continuum} + \text{resonance ($f_2$)}$.
We consider the resonances $f_0 (500)$ and $f_2 (1270)$ in our analysis.
Another scalar meson $f_0 (980)$ is neglected because there is 
no clear signal of $f_0 (980)$ in the differential cross-section data 
of Belle, although its effects are seen in the total cross section. 
Furthermore, its theoretical decay-constant estimate is not available 
by considering that it is a tetra-quark state, 
which is likely to be the $f_0 (980)$ configuration \cite{f0-4q}.
The resonance terms contain resonance masses, total decay widths,
two-pion coupling constants, and decay constants, which are
taken from the particle-data-group tables and theoretical articles.
The continuum terms are expressed by the momentum fraction carried 
by quarks and antiquarks in the pion $M_{2(q)}^{\pi^0}$ and
the overall timelike form factor
$F^{\,\pi}_q (W^2) = 1 / [ 1 + (W^2-4 m_\pi^2)/\Lambda^2 ]^{n-1}$
with the constituent-counting factor $n=2$ \cite{counting}.
The cutoff parameter $\Lambda$ is one of the parameters
in the $\chi^2$ analysis.
Because of the page limitation, the details of $\widetilde B_{nl}(W^2)$
and our parametrization are not explained here, and they
should be found in the original paper \cite{gdas-kst-2017}.

\begin{figure}[t!]
\begin{minipage}{\textwidth}
\begin{tabular}{lcl}
\begin{minipage}[c]{0.48\textwidth}
     \includegraphics[width=5.3cm]{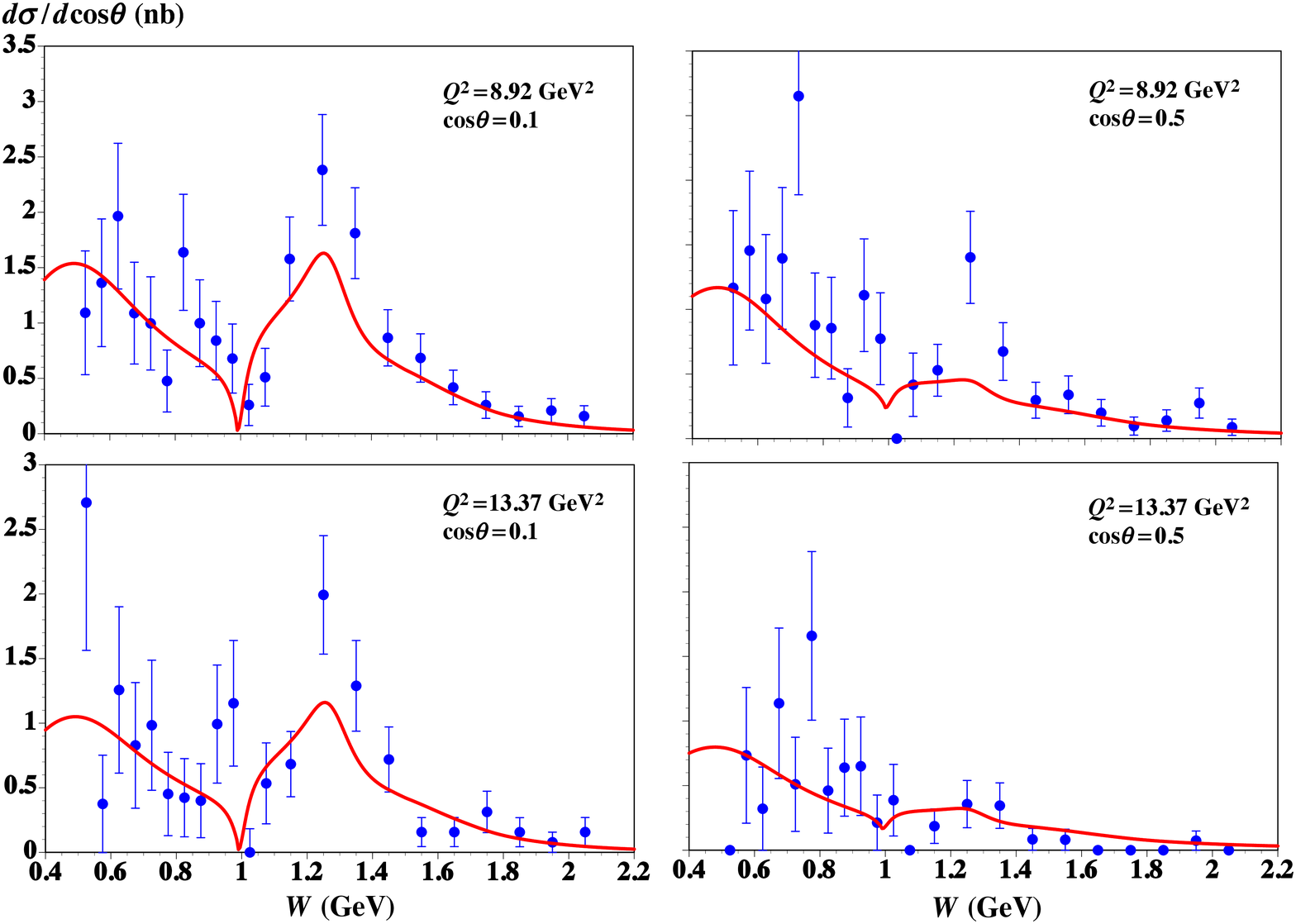}
\end{minipage} 
\hspace{0.30cm}
\begin{minipage}[c]{0.48\textwidth}
     \includegraphics[width=5.3cm]{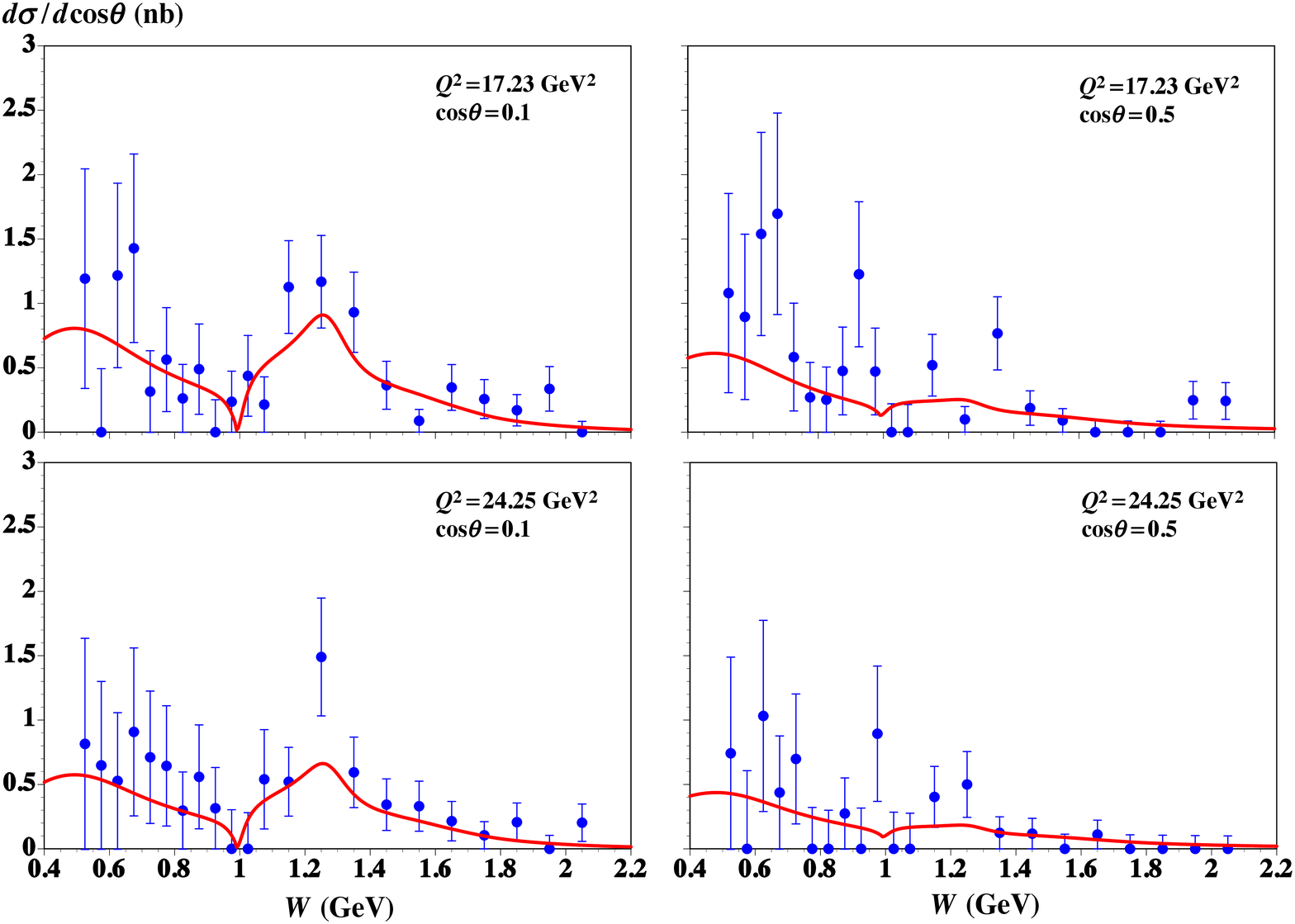}
\end{minipage}
\end{tabular}
\end{minipage}
\caption{Comparison with differential cross-section data
of the Belle collaboration
at $Q^2=8.92$, 13.37, 17.23, and 24.25 GeV$^2$
with $\cos\theta =0.1$ and 0.5 \cite{gdas-kst-2017}.}
\label{fig:comparison-belle}
\vspace{-0.3cm}
\end{figure}

We analyzed the Belle experimental data on the differential
cross section for $\gamma^* \gamma \to \pi^0 \pi^0$,
and the optimum GDAs are determined by the $\chi^2$ analysis.
The comparison with some Belle data are shown in 
Fig.\,\ref{fig:comparison-belle}, where the $Q^2$ values are
$Q^2=8.92$, 13.37, 17.23, and 24.25 GeV$^2$ and the scattering
angles are $\cos\theta =0.1$ and 0.5.
The smaller $Q^2$ data are not included in our analysis by 
considering the factorization condition $Q^2 \gg W^2$.
The curves indicate our theoretical results, and
they explain the data reasonably well. There are peaks
in the $W=1.3$ GeV region and it comes from the $f_2 (1270)$ resonance,
and $f_0 (500)$ affects the cross section in the small-$W$ region.
There is an overall contribution from the GDA continuum
in the whole-$W$ range of Fig.\,\ref{fig:comparison-belle}.

\begin{figure}[b!]
\vspace{-0.00cm}
\begin{minipage}{\textwidth}
\begin{tabular}{lc}
\hspace{-0.30cm}
\begin{minipage}[c]{0.47\textwidth}
   \vspace{-0.2cm}
   \begin{center}
    \includegraphics[width=4.6cm]{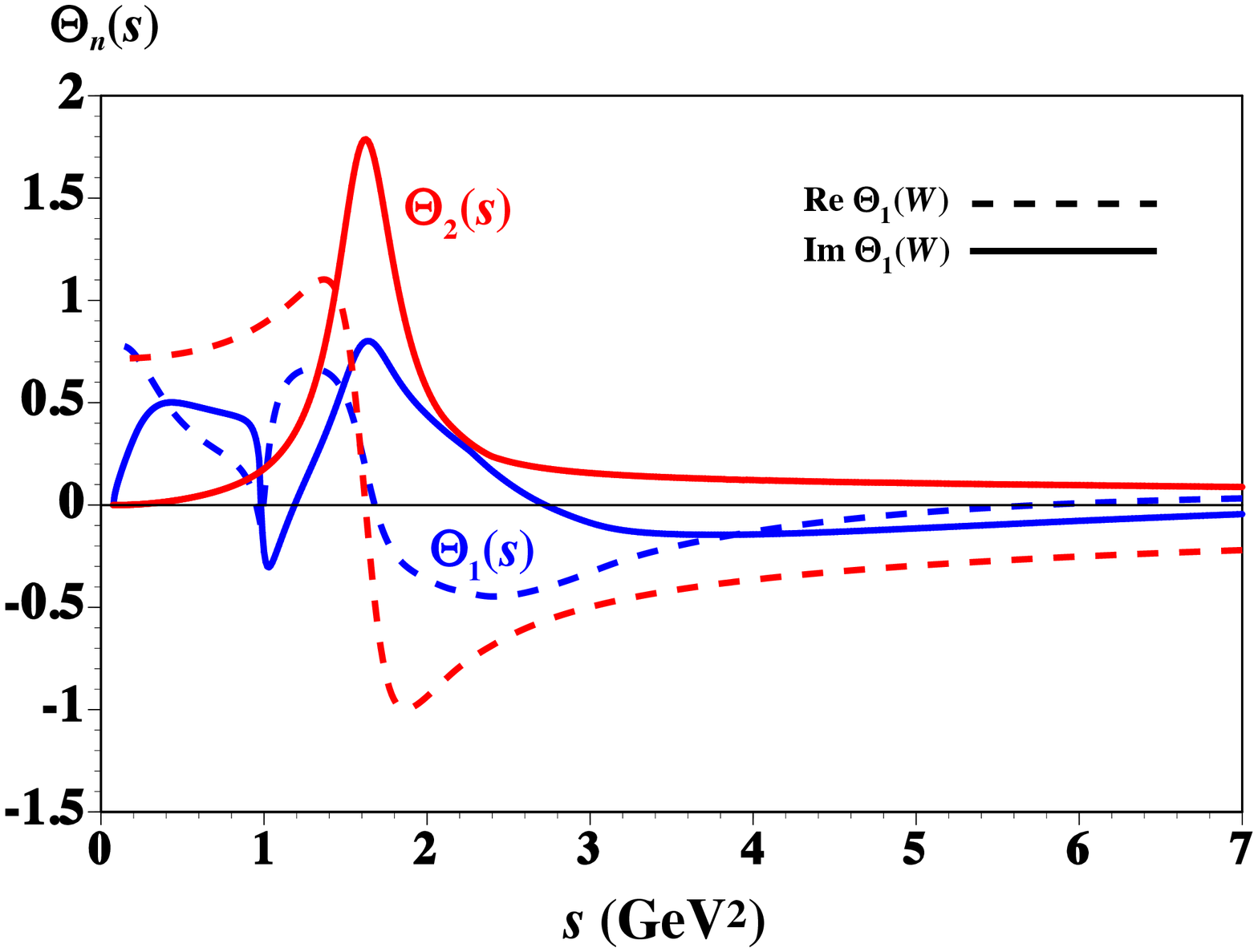}
   \end{center}
\vspace{-0.40cm}
\caption{Timelike gravitational form factors for $\pi$ \cite{gdas-kst-2017}.}
\label{fig:theta12reim-time}
\vspace{-0.4cm}
\end{minipage} 
\hspace{0.5cm}
\begin{minipage}[c]{0.47\textwidth}
    \vspace{-0.2cm}
   \begin{center}
    \includegraphics[width=4.6cm]{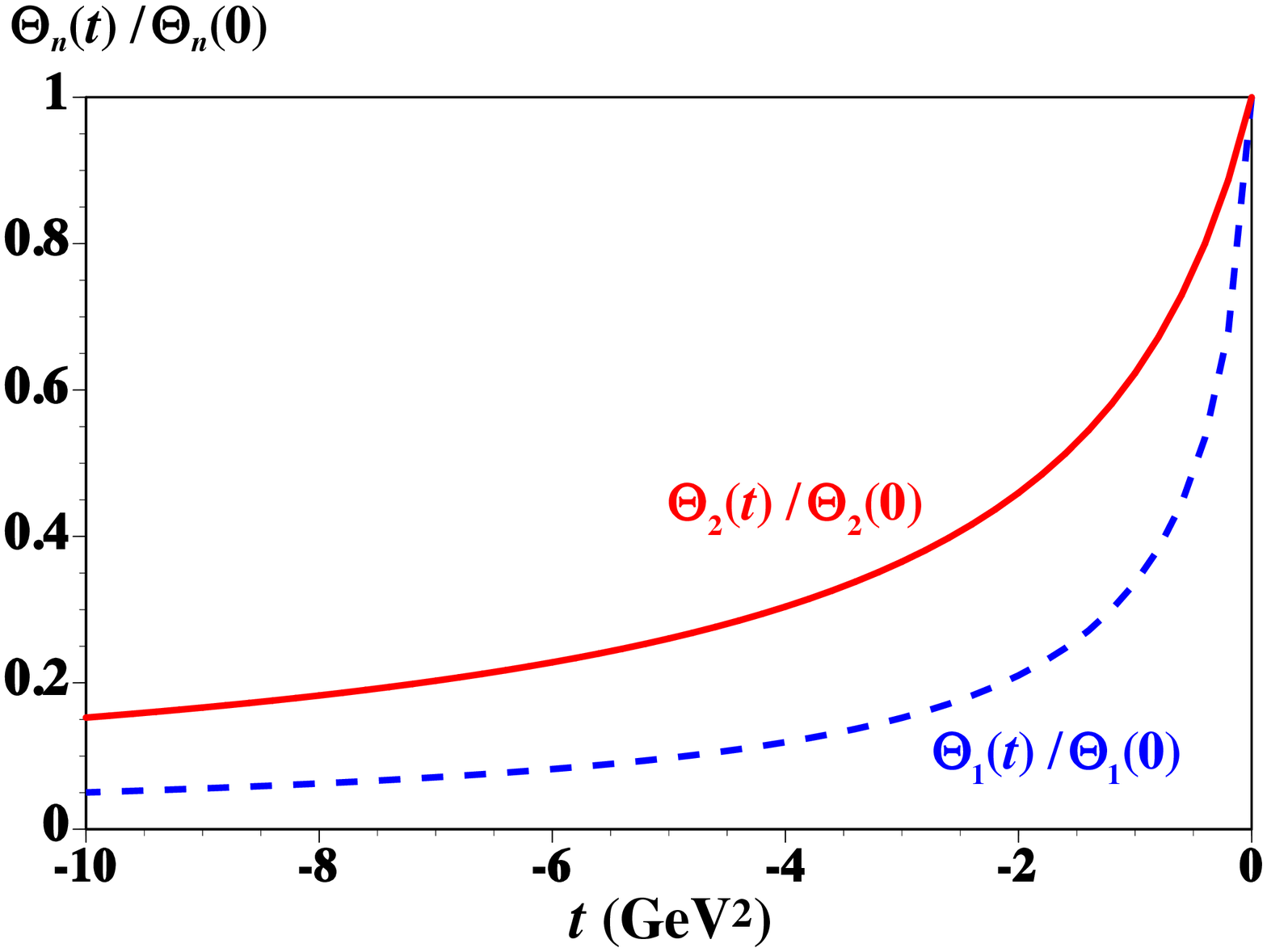}
   \end{center}
\vspace{-0.40cm}
\caption{Spacelike gravitational form factors for $\pi$ \cite{gdas-kst-2017}.}
\label{fig:theta12-space}
\vspace{-0.4cm}   
\end{minipage}
\end{tabular}
\vspace{0.20cm}
\end{minipage}
\end{figure}

Using the form-factor definition and the GDAs
in Eqs.\,(\ref{eqn:emt-ffs-timelike-0}) and
(\ref{eqn:gda-parametrization}), we obtain
the form factors expressed by the $W$-dependent functions
of the GDAs as
\begin{align}
\Theta_{1, q} (s) 
= -\frac{3}{5} \widetilde B_{10} (W^2) 
+ \frac{3}{10} \widetilde B_{12} (W^2) ,
\ \ \ 
\Theta_{2, q} (s) 
= \frac{9}{10 \, \beta^2} \widetilde B_{12} (W^2) .
\label{eqn:emt-ffs-gdas}
\end{align}
Then, the total timelike gravitational form factors of the pion
are obtained by adding them as
$ \Theta_{n} (s) = \sum_{i=q} \Theta_{n, i} (s)$ where $n=1$ or 2,
and they are shown in Fig.\,\ref{fig:theta12reim-time}.
The D-wave term contributes to the form factor $\Theta_2$,
which shows the resonance behavior at the $f_2$ mass.
The function $\Theta_1$ has more complicated $W$ dependence
due to the additional S-wave term. Since the imaginary parts 
of the form factors are determined in our analysis, they
are used for calculating the spacelike gravitational 
form factors by using the dispersion relation.
The obtained spacelike form factors are shown 
in Fig.\,\ref{fig:theta12-space}.
There are significant differences in the $t$ dependence between
the two form factors due to the additional S-wave term.

For understanding the meaning of the form factors
$\Theta_1$ and $\Theta_2$,
we may define the static energy-momentum tensor as \cite{static-form} 
$ T^{\mu\nu}_q (\vec r \,) = \int d^3 q / [(2\pi)^3 \, 2E] 
e^{i \vec q \cdot \vec r} \\
\left\langle \pi^0 (p') \! \left| 
T^{\mu\nu}_q (0) \, \right| \! \pi^0 (p) \right\rangle $,
where $E$ is the pion energy $E=\sqrt{m_\pi^2 +\vec q^{\ 2}/4}$.
The $\mu\nu = ij$ ($i,\, j =1,\, 2,\, 3$) components
are expressed by the pressure $p(r)$ and shear force $s(r)$ as
$ T^{\, ij}_q (\vec r \,) = p_q (r) \, \delta_{ij} 
    + s_q (r) ( r_i r_j / r^2 - \delta_{ij} /3 )$.
The term $T^{\, ij}_q (\vec r \,)$ is expressed only by $\Theta_1$,
so that $\Theta_1$ indicates pressure and shear-force distributions.
We may call it as mechanical (pressure, shear-force) form factor.
On the other hand, the $\mu\nu = 00$ component satisfies relation
$ \int d^3 r \, T^{00}_q (\vec r \,) = m_\pi \Theta_{2,q} (0)$.
It means that $\Theta_2$ shows the gravitational mass (or energy) 
distribution in the pion.
At finite $t$, $\Theta_1$ also contributes to this distribution.

\begin{wrapfigure}[11]{r}{0.46\textwidth}
   \vspace{-0.7cm}
   \begin{center}
    \includegraphics[width=4.6cm]{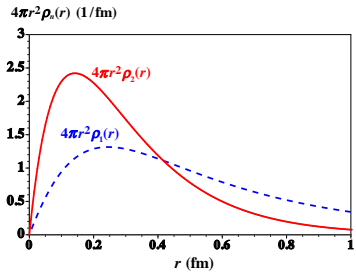}
   \end{center}
\vspace{-0.45cm}
\caption{Mass and mechanical densities \cite{gdas-kst-2017}.}
\label{fig:rho12}
\vspace{-0.7cm}
\end{wrapfigure}

By taking the Fourier transforms of the spacelike form factors,
the space-coordinate densities are obtained as shown 
in Fig.\,\ref{fig:rho12}.
The mechanical distribution $\rho_1 (r)$ extends to the larger-$r$ region
than the gravitational-mass distribution $\rho_2 (r)$.
From the form factors or densities, 
the root-mean-square radii are calculated as:
$\sqrt {\langle r^2 \rangle _{\text{mass}}} = 0.69 \, \text{fm}$ and
$\sqrt {\langle r^2 \rangle _{\text{mech}}} = 1.45 \, \text{fm}$,
which are gravitational-mass and mechanical radii, respectively.
There are some ambiguities in our results due to phase-factor
assignment. If we consider such ambiguities, we obtain the radius
ranges as \cite{gdas-kst-2017}:
\begin{align}
\sqrt {\langle r^2 \rangle _{\text{mass}}} 
    =  0.56 \sim 0.69 \, \text{fm}, \ \ \ 
\sqrt {\langle r^2 \rangle _{\text{mech}}} 
    = 1.45 \sim 1.56 \, \text{fm} .
\label{eqn:g-radii-pion-range}
\end{align}
This should be the first result on the gravitational radii 
from the actual analysis of experimental measurements.
The charge radius of the pion has been already measured as
$\sqrt {\langle r^2 \rangle _{\text{charge}}} =0.672 \pm 0.008$ fm
\cite{pdg-2016}.
It is our interesting finding that the gravitational-mass
radius is similar or slightly smaller than the charge radius
and that the mechanical radius is larger.

\vspace{-0.10cm}



\begin{thebibliography}{}
\vspace{-0.20cm}
\bibitem{gpds-gdas}
M.~Diehl, T.~Gousset, B.~Pire, and O.~Teryaev,
  Phys.\ Rev.\ Lett.\  {\bf 81}, 1782 (1998);
For a review, see
M.~Diehl, Phys. Rept.  {\bf 388}, 41 (2003).
\bibitem{gdas-kk-2014}
  H.~Kawamura and S.~Kumano, Phys. Rev. D {\bf 89}, 054007 (2014).
\bibitem{gdas-kst-2017}
  S. Kumano, Qin-Tao Song, and O. V. Teryaev, arXiv:1711.08088,
  Phys. Rev. D in press.
\bibitem{Masuda:2015yoh}
  M.~Masuda {\it et al.} (Belle Collaboration),
  Phys. Rev. D {\bf 93}, 032003 (2016).
\bibitem{f0-4q}
  S. Kumano, V. R. Pandharipande, Phys. Rev. D {\bf 38}, 146 (1988);
  F. E. Close, N. Isgur, and S. Kumano, Nucl. Phys. B {\bf 389}, 513 (1993);
  T. Sekihara and S. Kumano, Phys. Rev. D {\bf 92}, 034010 (2015).
\bibitem{counting}
  H. Kawamura, S. Kumano, and T. Sekihara, Phys. Rev. D {\bf 88}, 034010 (2013).
  W.-C. Chang, S. Kumano, and T. Sekihara, Phys. Rev. D {\bf 93}, 034006 (2016).
\bibitem{static-form} 
   M. V. Polyakov, Phys. Lett. B {\bf 555}, 57 (2003).
\bibitem{pdg-2016}
  C. Patrignani {\it et al.} (Particle Data Group), 
      Chin. Phys. C {\bf 40}, 100001 (2016). 
\end{thebibliography}
\end{document}